\documentclass[aps,prl,preprint,letterpaper,showpacs,superscriptaddress]{revtex4-1}
\usepackage{natbib}
\usepackage{graphicx}
\usepackage{amsmath}
\usepackage{amsthm}
\usepackage{amsbsy}
\usepackage{amsfonts}
\usepackage{mathrsfs}
\usepackage{subfigure}
\usepackage{pslatex}
\newfont{\tensy}{cmsy10}

\newcommand{\cref}[1]{Chap.\ \ref{#1}}

\def\bea{\begin{eqnarray}}
\def\eea{\end{eqnarray}}
\def\ben{\begin{equation}}
\def\een{\end{equation}}

\newcommand{\I}{\mathrm{i}}
\begin{document}
%%%%%%%%%%%%%%%%%%%%%%%%%%%%%%%%%%%%%%%%%%%%%%%%%%%%%%%%%%%%%%%%%%%%%%%%%%%%%%%
%%%%%%%%%%%%%%%%%%%%%%%%%%%%%%%%%%%%%%%%%%%%%%%%%%%%%%%%%%%%%%%%%%%%%%%%%%%%%%%
%%%%%%%%%%%%%%%%%%%%%%%%%%%%%%%%%%%%%%%%%%%%%%%%%%%%%%%%%%%%%%%%%%%%%%%%%%%%%%%

%%%%%%%%%%%%%%%%%%%%%%%%%%%%%%%%%%%%%%%%%%%%%%%%%%%%%%%%%%%%%%%%%%%%%%%%%%%%%%%
% Title, author, affiliations.
\title{Controlling the Dynamics of Many-Electron Systems from First Principles:
A Marriage of Optimal Control and Time-Dependent Density-Functional Theory}

\author{A. Castro}
\affiliation{
Institute for Biocomputation and Physics of Complex Systems (BIFI), University of Zaragoza, E-50018 Zaragoza, Spain
}
\author{J. Werschnik}
\affiliation{
JENOPTIK Laser, Optik, Systeme GmbH, Jena, Germany
}
\author{E. K. U. Gross}
\affiliation{
Max-Planck-Institut f{\"{u}}r Mikrostrukturphysik, Weinberg 2, D-06120 Halle, Germany
}
\date{\today}
%%%%%%%%%%%%%%%%%%%%%%%%%%%%%%%%%%%%%%%%%%%%%%%%%%%%%%%%%%%%%%%%%%%%%%%%%%%%%%%

%%%%%%%%%%%%%%%%%%%%%%%%%%%%%%%%%%%%%%%%%%%%%%%%%%%%%%%%%%%%%%%%%%%%%%%%%%%%%%%
% ABSTRACT
\begin{abstract}
  Quantum Optimal Control Theory (QOCT) provides the necessary tools
  to theoretically design driving fields capable of controlling a
  quantum system towards a given state or along a prescribed path in
  Hilbert space. This theory must be complemented with a suitable
  model for describing the dynamics of the quantum system. Here, we
  are concerned with many electron systems (atoms, molecules, quantum
  dots, etc) irradiated with laser pulses. The full solution of the
  many electron Schr{\"{o}}dinger equation is not feasible in general,
  and therefore, if we aim to an {\emph{ab initio}} description, a
  suitable choice is time-dependent density-functional theory
  (TDDFT). In this work, we establish the equations that combine TDDFT
  with QOCT, and demonstrate their numerical feasibility with examples.
\end{abstract}
%%%%%%%%%%%%%%%%%%%%%%%%%%%%%%%%%%%%%%%%%%%%%%%%%%%%%%%%%%%%%%%%%%%%%%%%%%%%%%%

%%%%%%%%%%%%%%%%%%%%%%%%%%%%%%%%%%%%%%%%%%%%%%%%%%%%%%%%%%%%%%%%%%%%%%%%%%%%%%%
% MAKETITLE
% Klassifizierung ueber PACS-Schema rausfinden,
%siehe: http://publish.aps.org/infoauth.html
\maketitle
%%%%%%%%%%%%%%%%%%%%%%%%%%%%%%%%%%%%%%%%%%%%%%%%%%%%%%%%%%%%%%%%%%%%%%%%%%%%%%%

%%%%%%%%%%%%%%%%%%%%%%%%%%%%%%%%%%%%%%%%%%%%%%%%%%%%%%%%%%%%%%%%%%%%%%%%%%%%%%%
% TEXT BODY

The quest for systems able to perform quantum
computing~\cite{Tesch2001633, *PhysRevLett.89.188301}, the synthesis
of design-molecules by laser-induced chemical
reactions~\cite{laarmann:201101}, or the control of electron currents
in molecular switches using light~\cite{geppert-2004} may benefit from
the recent advances in the field of design and synthesis of laser
pulses specially tailored to perform a specific
task~\cite{weiner-2000}.
%A number
%of experiments have demonstrated the possibility of tailoring this
%type of laser pulses in order to produce a desired reaction: The
The laser pulse creation and shaping techniques have improved impressively
over the last decades, and thus the area of experimental optimal
control has become a well established field.

Such pulses can also be theoretically derived with the help of quantum
optimal control theory~\cite{shapiro-2003, *rice-2000,
  *werschnik-2007, [{For recent developments, see the dedicated volume
      edited by }] [{}]rabitz-2009, phdbook} (QOCT).  This theory is
rather general in scope, and its basic formulation makes no assumption
on the nature and modelling of the quantum system on which the pulse
is applied.  In practice, the solution of the QOCT equations requires
multiple propagations, both forwards and backwards, for the system
under study. Since these propagations are in general unfeasible for
many-particle systems, few-level simplifications and models are
typically postulated when handling the QOCT equations. Unfortunately,
these simplifications are not always accurate enough: strong pulses
naturally involve many levels, and normally perturbative treatments
are not useful. Non linear laser-matter interaction must sometimes be
described \emph{ab initio}.

In this work, we are concerned with many-electrons systems (atoms,
molecules, quantum dots\dots) irradiated with femtosecond (or
attosecond) pulses, with intensities typically ranging from 10$^{11}$
to 10$^{15}$ Wcm$^{-2}$ (a non-linear regime that nevertheless allows
for a non-relativistic treatment). This may lead to a number of
interesting phenomena, e.g. above-threshold or tunnel ionization, bond
hardening or softening, high harmonic generation, photo-isomerization,
photo-fragmentation, Coulomb explosion, etc~\cite{protopapas-1997,
  *brabec-2000, *scrinzi-2006}.  In order to describe this type of
processes from first principles, time-dependent density-functional
theory~\cite{RG84, *marques-book-2006} (TDDFT) has emerged as a viable
alternative to more computationally expensive approaches based on the
wave function.

In TDDFT, the system of interacting electrons is substituted by a
proxy system of non-interacting electrons -- the ``Kohn-Sham'' (KS)
system, which is computationally much less demanding.  The theory
guarantees the identity of the electronic densities of the two
systems, and the existence of a density functional for each possible
observable, thus allowing (in principle) the computation of any
property without ever dealing with the many-body wave function. The
theory is however hindered by the lack of knowledge of the precise
external potential seen by the auxiliary non-interacting system (the
so-called ``exchange and correlation'' potential, which is a
functional of the density itself, has to be approximated), and, in
many cases, the precise form of the density functional that provides
the required observable. Fortunately, a number of valid approximations
for these density functionals have been developed over the years,
which have made of TDDFT a computationally efficient possibility to
describe many processes.
%%%%%%%%%% 
% WARNING:
% IN ORDER TO SHORTEN THE PAPER, I REMOVE THIS.
%%%%%%%%%%
%The use of the density as a basic variable also implies
%that we will not have any good approximation for the wave function,
%which leads to some inconveniences for the QOCT scheme, to be
%discussed below.

We are thus led to the necessity of inscribing TDDFT into the general
QOCT framework. We will lay down and discuss the equations that result
when TDDFT is used to model the system. Then, in order to demonstrate
the computational feasibility, we present one sample calculation: a 2D
two-electron system being optimally driven between two potential
wells.

%\emph{
%So far, research in this area is mainly focussed on one-particle
%systems. A generalization to many-particle systems is in principle
%straightforward. In practice, however, the iterative calculation of
%the optimal laser field requires solving the time-dependent many-body
%Schr\"odinger equation many times \cite{ZBR98,OTR2004}. This limits
%the applicabilty to systems with a very small number of particles
%\cite{phdbook}.  The best way, in our opinion, to overcome this
%limitation is the combination of OCT with time-dependent density
%functional theory (TDDFT) \cite{RG84,GDP96,BWG2005}. This letter
%presents the formalism, an example calculation, and the problems that
%emerge when combining TDDFT and OCT.
%}

In the spirit of TDDFT, we substitute the problem of formulating QOCT
in terms of the real interacting system, by the formulation of the
optimization problem for the non-interacting system of electrons.  The
equations of motion for the single-particle orbitals of this system,
also known as time-dependent KS (TDKS) equations, are:
\begin{align}
\nonumber
\I \frac{\partial \varphi_i}{\partial t}(\vec{r}\sigma,t) = \langle \vec{r}\sigma \vert
\hat{H}_{\rm KS}[n_{\tau\omega}(t),u,t] \vert \varphi_i(t)\rangle = 
\\
\nonumber
-\frac{1}{2}\nabla^2 \varphi_i(\vec{r}\sigma,t) + 
\langle \vec{r}\sigma \vert \hat{V}_0\vert\varphi_i(t)\rangle  + v_{\rm H}[n(t)](\vec{r})\varphi_i(\vec{r}\sigma,t) + 
\\
\langle \vec{r}\sigma \vert \hat{V}_{\rm xc}[n_{\tau\omega}(t)]\vert \varphi_i(t)\rangle + 
\langle \vec{r}\sigma \vert \hat{V}_{\rm ext}[u]\vert
\varphi_i(t)\rangle\,,
\end{align}
\begin{equation}
n_{\tau\omega}(\vec{r},t) = \sum_{i=1}^N
\varphi^*_i(\vec{r}\tau, t)\varphi_i(\vec{r}\omega,t)\,,
\end{equation}
\begin{equation}
n(\vec{r},t) = \sum_{\sigma} n_{\sigma\sigma}(\vec{r},t)\,,
\end{equation}
for $i=1,\dots,N$ orbitals.  The greek indexes
$\sigma,\tau,\omega\dots$ run over the two spin configurations, up and
down. The densities are, by construction, equal to that of the
\emph{real}, interacting system of electrons. $\hat{V}_0$ represents
the internal, time independent fields -- usually a nuclear Coulomb
potential $V_{\rm n}(\vec{r})$, and may include as well a spin-orbit
coupling term of the form $\vec{\sigma} \cdot \vec{\nabla}V_{\rm n}
\times \hat{\vec{p}}$ (where $\vec{\sigma}$ is the vector of Pauli
matrices).  The term $v_H[n(t)](\vec{r}) = \int\!\!  {\rm
  d}^3r'\;\frac{n(\vec{r}',t)}{\vert\vec{r}-\vec{r}'\vert}$ is the
Hartree potential, and $\hat{V}_{\rm xc}[n_{\tau\omega}]$ is the
exchange and correlation potential operator, whose action is given by:
\begin{equation}
\langle \vec{r}\sigma\vert \hat{V}_{\rm xc}[n_{\tau\omega}(t)]\vert \varphi_i(t)\rangle = 
\sum_\beta v_{\rm xc}^{\sigma\beta}[n_{\tau\omega}(t)](\vec{r})\varphi_i(\vec{r}\beta,t)\,.
\end{equation}
Note that, for the sake of generality, we allow a spin-resolved
exchange and correlation potential, that depends on the four spin
components $n_{\tau\delta}$ (in many situations more restricted
dependences are assumed). However, we do assume here an
\emph{adiabatic} approximation, i.e. $v_{\rm xc}$ at each time $t$ is
a functional of the densities at that time, $n_{\tau\omega}(t)$. This
restriction is non-essential for the derivations that follow, but the
use of non-adiabatic functionals is very scarce, and adiabatic
approximations will result in simpler equations.

The last potential term, $\hat{V}_{\rm ext}$, is the external
time-dependent potential, which is determined by a ``control'' $u$.  In
a typical case, this external potential is the electric pulse created
by a laser source in the dipole approximation, and $u$ is the real
time-dependent function that determines its temporal shape (in this
case, $\langle \vec{r}\sigma \vert \hat{V}_{\rm ext}[u]\vert
\varphi_i(t)\rangle = u(t)\vec{r}\cdot\vec{p}\varphi_i(\vec{r}\sigma,
t)$, where $\hat{p}$ is the polarization vector of the pulse). We
write it however in general operator form, since it can be a 2x2
matrix which may include both a time-dependent electric field as well
as a Zeeman-coupled magnetic field~\cite{PhysRevLett.105.097205}.  The
mathematical nature of $u$ may also be diverse: it may not be a
time-dependent function, but a set of $N$ parameters that determine
the precise form of the electric field.

If we group the $N$ single particle states into a vector
$\underline{\varphi}(t)$, we can rewrite the TDKS equations in a matrix form:
\begin{equation}
\I\underline{\dot{\varphi}}(t) = \underline{\underline{\hat{H}}}[n_{\tau\omega}(t),u,t]
\underline{\varphi}(t)\,,
\end{equation}
where $\underline{\underline{\hat{H}}}[n_{\tau\omega}(t),u,t]=\hat{H}_{\rm
  KS}[n_{\tau\omega}(t),u,t]\underline{\underline{I}}_N$ and
$\underline{\underline{I}}_N$ is the $N$-dimensional unit matrix.
With this notation we stress the fact that we have only \emph{one}
dynamical system -- and not $N$ independent ones, since all
$\varphi_i$ are coupled. This coupling, however, comes solely through
the density, since the Hamilton matrix is diagonal.

The specification of the value of the control $u$, together with the
initial conditions, determines the solution orbitals: $u \rightarrow
\underline{\varphi}[u]$.  Our task is now the following: we wish to
find an external field -- in the language of OCT, a control $u$ --
that induces some given behaviour of the system, which can be
mathematically formulated by stating that the induced dynamics
maximizes some target functional $F$. Since we are using TDDFT, this
functional will be defined in terms of the KS orbitals, and will
possibly depend explicitly on the control $u$:
\begin{equation}
F = F[\underline{\varphi},u]\,.
\end{equation}
Since the KS orbitals depend on $u$ as well, the goal of QOCT can be
formulated as finding the maximum of the function:
\begin{equation}
G[u] = F[\underline{\varphi}[u],u]\,.
\end{equation}
In the most general case, the functional $F$ depends on $\underline{\varphi}$ at
all times during the process (we have a ``time-dependent target''). In
many cases, however, the goal is the achievement of some target at a
given time $T$ that determines the end of the propagation interval (we
then have a ``static target''). In both cases, the determination of
the value of the function $G$ is obtained by performing the
propagation of the system with the field determined by the control
$u$.

There are many optimization algorithms capable of maximizing functions
utilizing solely the knowledge of the function values (``gradient-free
algorithms''). We have recently employed one of them in this
context~\cite{castro-2009}. However, QOCT provides the solution to the
problem of computing the gradient of $G$ -- or, properly speaking, the
functional derivative if $u$ is a function. The non-linear dependence
of the Hamiltonian with the density slightly complicates the
derivation, but we sketch the key steps: First, we must note that
searching for a maximum of $G$ is equivalent to a constrained search
for $F$ -- constrained by the fact that the $\underline{\varphi}$
orbitals must fulfill the TDKS equations. In order to do
so, we introduce a new set of orbitals $\underline{\chi}$ that act as
Lagrange multipliers, and define a new functional $J$ by adding a
Lagrangian term $L$ to $F$:
\begin{equation}
J[\underline{\varphi},\underline{\chi},u] = F[\underline{\varphi},u] 
+ L[\underline{\varphi},\underline{\chi},u]\,,
\end{equation}
\begin{equation}
L[\underline{\varphi},\underline{\chi},u] = -2 \sum_{j=1}^N{\rm Re}\left[
\int_0^T\!\!\!\!\! {\rm d}t\; \langle \chi_j(t)\vert
\I\frac{\rm d}{{\rm d}t} + \hat{H}_{\rm KS}[n_{\tau\omega}(t),u,t]\vert\varphi_j(t)\rangle
\right]\,.
\end{equation}
Setting the functional derivatives of $J$ with respect to the $\chi$
orbitals to zero, we retrieve the TDKS equations. In an analogous
manner, we obtain a set of solution $\underline{\chi}[u]$ orbitals by taking
functional derivatives with respect to $\underline{\varphi}$:
%and therefore, the
%$\underline{\varphi} \to \underline{\varphi}_u$ mapping:
%\begin{equation}
%\frac{\delta J}{\delta \chi_j} = 0 \Rightarrow \varphi_{j} \to \varphi_{j,u}\,.
%\end{equation}
\begin{equation}
\nonumber
\frac{\delta J}{\delta \underline{\varphi}^*} = 0 \Rightarrow %\chi_{j} \to \chi_{j,u}\,.
\end{equation}
\begin{eqnarray}
\label{eq:backwards}
\I\underline{\dot{\chi}}(t) & = & \left[
\underline{\underline{\hat{H}}}_{\rm KS}[n_{\tau\sigma}[u](t),u,t] + \underline{\underline{\hat{K}}}[\underline{\varphi}[u](t)]
\right]
\underline{\chi}(t)
-\I\frac{\delta F}{\delta \underline{\varphi}^*}
%[\underline{\varphi}_u(t);u]
\,,
\\
\label{eq:backwards-finalvalue}
\underline{\chi}(T) & = & \underline{0}\,.
\end{eqnarray}
The non-diagonal operator matrix
$\underline{\underline{\hat{K}}}[\underline{\varphi}[u](t)]$ is defined by
the operators:
\begin{align}
\nonumber
\langle \vec{r}\sigma \vert \hat{K}_{ij}[\underline{\varphi}[u](t)]\vert \psi \rangle  =  
-2\I \sum_{\kappa} \varphi_i[u](\vec{r}\kappa,t) \times
\\
{\rm Im}\left[ \sum_{\alpha\beta}
\int {\rm d}^3r' \psi^*(\vec{r}'\alpha)f^{\alpha\beta,\sigma\kappa}_{\rm Hxc}[n_{\tau\omega}[u](t)](\vec{r},\vec{r}')\varphi_j[u](\vec{r}'\beta,t)
\right]\,,
\end{align}
\begin{equation}
f^{\alpha\beta,\gamma\delta}_{\rm Hxc}[n_{\tau\omega}](\vec{r},\vec{r}') = \frac{\delta_{\alpha\beta}\delta_{\gamma\delta}}{\vert\vec{r}-\vec{r}'\vert} 
+ \frac{\delta v^{\alpha\beta}_{\rm xc}[n_{\tau\omega}](\vec{r})}{\delta n_{\gamma\delta}(\vec{r}')}\,.
\end{equation}
If we now note that $G[u] =
J[\underline{\varphi}[u],\underline{\varphi}[u],u]$, we arrive to:
\begin{align}
\nonumber
\nabla_u G[u]  =  \nabla_u \left.F[\underline{\varphi},u]\right|_{\underline{\varphi}=\underline{\varphi}[u]} + 
\\
\label{eq:main-eq}
2 {\rm Im}\left[
\sum_{j=1}^N \int_0^T \!\!\!{\rm d}t\; \langle \chi_j[u](t)\vert \nabla_u \hat{V}_{\rm ext}[u](t)
\vert \varphi_j[u](t)\rangle
\right]
\end{align}

Several aspects of these equations deserve further discussion:

\emph{(1)} Eqs.~\ref{eq:backwards} and ~\ref{eq:backwards-finalvalue}
are a set of first-order differential equations, whose solution must
be obtained by \emph{backwards} propagation, since the boundary
condition, Eq.~\ref{eq:backwards-finalvalue} is given at the end of
the propagating interval, $T$. Note that this propagation depends on
the time-dependent KS orbitals $\underline{\varphi}[u]$. Therefore, the
numerical procedure to follow consists of a forward propagation to
obtain $\underline{\varphi}[u]$, followed by a backwards propagation to
obtain $\underline{\chi}[u]$.

\emph{(2)} These backwards equations are non-homogeneous, due to the
presence of the last term in Eq.~\ref{eq:backwards}, the functional
derivative of $F$ with respect to $\underline{\varphi}$ -- but see
point (4) below.

\emph{(3)} Often, the control target functional $F$ is split like:
  \begin{equation}
  F[\underline{\varphi},u] = J_1[\underline{\varphi}] + J_2[u]\,.
  \end{equation}
  $J_1$ codifies the actual purpose of the optimization, whereas $J_2$
  imposes a \emph{penalty} on the control function, in order to avoid,
  for example, the solution field to have unreasonable amplitudes.
  In the following, we will assume this division.

  \emph{(4)} The previous equations
  (\ref{eq:backwards}-\ref{eq:backwards-finalvalue}) refer to a general
  ``time-dependet target'' case, as mentioned above. In many cases of
  interest, the target functional $F$ takes a ``static'' form, which
  can be expressed as:
  \begin{equation}
  J_1[\underline{\varphi},u] = O[\underline{\varphi}(T),u]\,,
  \end{equation}
  for some functional $O$ whose argument is not the full evolution of
  the KS system, but only its value at the end of the propagation. In
  this case, the non-inhomogeneity in Eq.~\ref{eq:backwards} vanishes,
  and instead we obtain a different final-value condition:
  \begin{equation}
  \label{eq:final-value}
  \chi_i[u](\vec{r}\sigma,T) = \frac{\delta O[\underline{\varphi}[u],u]}{\delta \varphi^*_i(\vec{r}\sigma)}\,.
  \end{equation}

\emph{(5)}
The previous Eq.~\ref{eq:main-eq} assumes that $u$ is a set of $N$ parameters,
$u\in\mathbb{R}^N$ that determines the control function. If $u$ is
directly the control function, the gradient has to be substituted by a
functional derivative, and the result is:
\begin{equation}
\frac{\delta G}{\delta u(t)}  =    \left.\frac{\delta F[\underline{\varphi},u]}{\delta u(t)}
\right|_{\underline{\varphi}=\underline{\varphi}[u]} + 
%\\
%& &
2 {\rm Im}\left[
\sum_{j=1}^N \langle \chi_j[u](t)\vert \hat{D}
\vert \varphi_j[u](t)\rangle
\right]\,.
\end{equation}
We have assumed here that the external potential $\hat{v}_{\rm ext}$
is determined by the function $u$ by a linear relationship:
\begin{equation}
\hat{V}_{\rm ext}[u](t) = u(t)\hat{D}\,.
\end{equation}
This is the most usual case ($\hat{D}$ would be the dipole operator,
and $u(t)$ the amplitude of an electric field),
but of course it would be trivial to generalize this to other
possibilities.

The previous scheme permits therefore to control the KS
system. However, the goal is to control the \emph{real} system. In
principle, the target is given by some functional $\tilde{J}_1[\Psi]$
that depends on the \emph{real} many-electron wave function of the
interacting system. This object is not provided by TDDFT, that only
provides the density $n$. Therefore, the ideal situation would be
that in which $\tilde{J}_1$ depends on $\Psi$ only through the density
$n$, $\tilde{J}_1=\tilde{J}_1[n]$. In this manner, optimizing for the KS
system is strictly equivalent to optimizing for the real one.  For
example, this holds if $\tilde{J}_1$ is given by the expectation value
of some one-body local operator $\hat{A}$:
\begin{equation}
\tilde{J}_1[\Psi] = \langle \Psi(T)\vert\hat{A}\vert\Psi(T)\rangle = 
\int {\rm d}^3r\; n(\vec{r},T)a(\vec{r})\,,
\end{equation}
where $\hat{A} = \sum_{i=1}^N a(\hat{\vec{r}}_i)$. In this case,
Eq.~\ref{eq:final-value} is simply:
\begin{equation}
\chi_i[u](\vec{r}\sigma,T) = a(\vec{r})\varphi_i[u](\vec{r}\sigma,T)\,.
\end{equation}
This will be the kind of target that we will be using in the example
below. Note that TDDFT ensures that all observables are functionals of
the density, and therefore \emph{in principle} one could always write
such functionals (we will provide another example in a forthcoming
publication, in which the target is the high harmonic generation
spectrum, which is also an explicit functional of the density).

\begin{figure}[t]
%\setlength{\unitlength}{\columnwidth}
%\setlength{\fboxsep}{0in}
%\framebox{
%\begin{picture}(1.00,0.70)
%\put(0.00,0.280){\framebox{\includegraphics[width=0.6\unitlength]{potential.eps}}}
%\put(0.60,0.42){\framebox{\includegraphics[width=0.40\unitlength]{wf1.eps}}}
%\put(0.60,0.14){\framebox{\includegraphics[width=0.40\unitlength]{wf2.eps}}}
%\put(0.00,0.00){\framebox{\includegraphics[width=0.40\unitlength]{opt-electric-field.eps}}}
%\end{picture}
\includegraphics[width=\columnwidth]{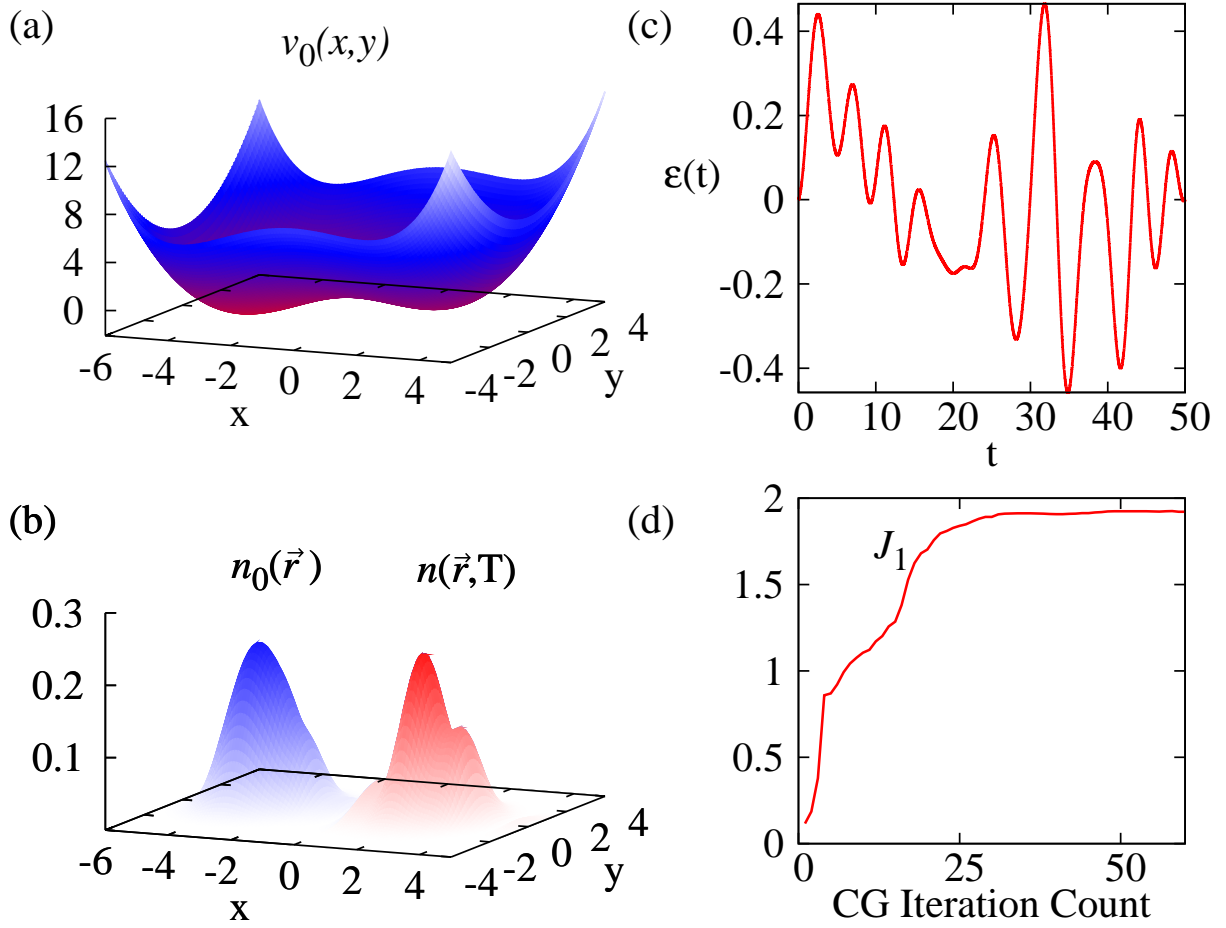}
%}
\caption{
\label{fig:1}
(a) External potential defining a model for double quantum dot. (b)
Density of the initial, ground state (blue) and final, propagated
density (red). (c) Optimized electric field for the charge-transfer
process described in the text. (d) Convergence history of the
conjugate gradient algorithm. All magnitudes are given in effective
atomic units.}
\end{figure}

%\begin{figure}[t]
%\centerline{\includegraphics[width=0.8\columnwidth]{opt-electric-field.eps}}
%\caption{
%\label{fig:2}
%Optimized electric field for the charge-transfer process described in the text.}
%\end{figure}

Unfortunately, in some cases we still need to find the explicit
density functional dependence in many cases of interest. For example,
a very common control goal is the transition from an initial state to
a target state. In other words, the control operator $\hat{A}$ is the
projection operator onto the target state $\hat{A}=\vert\Psi_{\rm
  target}\rangle\langle\Psi_{\rm target}\vert$. We have no exact
manner to substitute, in this case, the $\tilde{J}_1$ functional by a
functional $J_1$ defined in terms of the density, or in terms of the
KS determinant. It can be approximated, however, by an expression in
the form:
\begin{equation}
J_1[\underline{\varphi}] = 
\vert\langle\underline{\varphi}(T)\vert \sum_I c_I \vert \varphi^I\rangle\vert^2\,,
\end{equation}
where $\varphi(T)$ is the TDKS determinant at time $T$, and we compute
its overlap with a linear combination of Slater determinants
$\varphi^I$, weighted with some coefficients $c_I$. These Slater
determinants would be composed of occupied and unoccupied ground state
KS orbitals, $\varphi^I = {\rm
  det}[\varphi^I_1,\dots,\varphi^I_N]$. In this case,
Eq.~\ref{eq:final-value} takes the form:
\begin{equation}
\chi_i[u](\vec{r}\sigma,T) = \sum_{IJ} \lambda_{IJ}(\vec{r}\sigma) 
\langle \underline{\varphi}(T) \vert \varphi^I \rangle
\langle \varphi^J \vert \underline{\varphi}(T) \rangle\,,
\end{equation}
\begin{equation}
\lambda_{IJ}(\vec{r}\sigma) = c_Ic_J^* {\rm Tr}\lbrace
({\bf M}^I)^{-1} {\bf A}^i_I(\vec{r}\sigma)
\rbrace\,,
\end{equation}
where ${\bf M}^I_{mn} = \langle \varphi_m \vert \varphi^I_n\rangle$ and 
${\bf A}^i_I(\vec{r}\sigma)_{mn} = \delta_{mi}\varphi^I_n(\vec{r}\sigma)$.

%A prototypical example would be the attempt to drive a system from its
%closed shell ground state to an excited state. Let us assume that this
%state is a single-particle transition (e.g. an almost pure HOMO-LUMO
%excitation). If, for simplicity, we consider a two-electron system,
%the obvious way to approximate the excitations with a linear
%combination of KS Slater determinants would be:
%$\frac{1}{\sqrt{2}}\left( {\rm
%    det}[\varphi_0^\uparrow,\varphi_1^\downarrow] + {\rm
%    det}[\varphi_0^\downarrow,\varphi_1^\uparrow]\right)$, where
%$\varphi_0$ and $\varphi_1$ are the ground state and first unoccupied
%KS spatial orbitals, respectively, and $\varphi_i^\uparrow$ and
%$\varphi_i^\downarrow$ denote the spin-orbital formed by the
%$\varphi_i$ spatial orbital times spin up and down, respectively. It
%is then easy to prove that the maximum possible overlap of the TDKS
%state with this state is $\frac{1}{2}$~\cite{burke-2005}. However,
%this does not imply that the optimal pulse obtained in this way would
%produce, for the real interacting system, a $\frac{1}{2}$ occupation
%of the real excited state. In fact, it turns out that this occupation
%maybe be substantially larger~\cite{phdbook}.

We have implemented the described TDDFT+QOCT formalims in the {\tt
  octopus} code~\cite{octopus-1, *octopus-2}. In the following, we
describe a simple example: the charge transfer between two neighboring
potential wells, considered as models for 2D quantum dots, such as the
ones created in semiconductor heterostructures. We consider a
two-electron system, trapped in an asymmetric double quantum dot well
modelled by a potential function given by (in the following, we
consider {\rm effective} atomic units):
\begin{equation}
v_0(x,y) = \frac{1}{64}x^4-\frac{1}{4}x^2+\frac{1}{32}x^3+\frac{1}{2}y^2.
\end{equation}
The potential landscape is depicted in Fig.~\ref{fig:1}(a). We then
solve the ground state KS equations for this system, by making use of
the local density approximation to the exchange and correlation
parameterized by Attacalite et al~\cite{attacalite-2002}. The ground
state density will be localized in the left well (see
Fig.~\ref{fig:1}(b): $n_{\rm gs}(\vec{r}) = 2\vert\varphi_0^{\rm
  gs}(\vec{r})\vert^2$, where $\varphi^{\rm gs}_0$ is the ground state
KS orbital.

We apply an electric field, polarized along the $x$ direction. Its
amplitude is parameterized by its Fourier coefficients:
\begin{equation}
\epsilon(t) = \sum_{n=1}^{M/2}\left[ a_n\sqrt{\frac{2}{T}}\cos\left(\frac{2\pi}{T}nt\right) +
b_n\sqrt{\frac{2}{T}}\sin\left(\frac{2\pi}{T}nt\right)
\right]\,.
\end{equation}
The $\lbrace a_n,b_n\rbrace$ coefficients are therefore the control
$u$ (although the constraint $\sum_{n=1}^{N/2}a_n = 0$ is enforced in
order to ensure $\epsilon(0)=\epsilon(T)=0$). Since our goal is to
transfer as much charge as possible from the left to the right well,
%and we observe that the density corresponding to the first unoccupied
%ground state KS orbital $\varphi_1^{\rm gs}$ is localized in the right
%well [Fig.~\ref{fig:1}(b)], we can formulate a target in the form:
we formulate a target in the form:
\begin{equation}
\label{eq:target-example}
%  F[\underline{\varphi};a,b] = \vert\langle\varphi(T)\vert\varphi_1\rangle\vert^2 
%  - \alpha\sum_{n=1}^{M/2}\left(a_n^2+b_n^2\right)\,.
  F[\underline{\varphi};a,b] = \int_{x>0}\!\!\!\!\!\! {\rm d}^3r\; n(\vec{r},T)
  - \alpha\sum_{n=1}^{M/2}\left(a_n^2+b_n^2\right)\,.
\end{equation}
In words, we wish to arrive to a state in which all the density is
localized in the $x>0$ region.  The last term of
Eq.~\ref{eq:target-example} corresponds to the penalty:
\begin{equation}
J_2[a,b] = -\alpha\int_0^T\!\!\!\!\!{\rm d}t\; \epsilon^2(t) = 
-\alpha\sum_{n=1}^{M/2}\left(a_n^2+b_n^2\right)\,,
\end{equation}
and it is introduced in order to prevent the solution field from
having too much intensity.

The solution field is shown in Fig.~\ref{fig:1}(c). We have employed a
standard conjugate gradients (CG) algorithm to perform the
optimization. After around 60 CG iterations ~\footnote{ We are brief
  on numerical details. The interested reader may find the detailed
  description of the code used in Refs.~\cite{octopus-1,*octopus-2},
  and more updated information can be obtained at {\tt
    http://www.tddft.org/programs/octopus/}.  The code itself is
  available at that web page. Finally, see EPAPS Document No. [x],
  where the input and output files of the run are available.}, the
control field is converged and we achieve a value of $1.92$ for $J_1$
-- the maximum is 2 (see convergence plot in Fig.~\ref{fig:1}).

In conclusion, we have shown how TDDFT can be combined with QOCT, and
we have demonstrated how the resulting equations are numerically
tractable. This provides a scheme to perform QOCT calculations from
first principles, in order to obtain tailored function-specific laser
pulses capable of controlling the electronic state of atoms,
molecules, or quantum dots. Most of the previous applications of QOCT
were targeted to control, with femto-second pulses, the motion of the
nuclear wave packet on one or few potential energy surfaces, which
typically happens on a time scale of hundreds of femtoseconds or
picoseconds. The approach developed in this Letter, on the other hand,
is particularly suited to control the motion of the electronic degrees
of freedom which is governed by the sub-femto-second time scale.  The
possibilities that are open thanks to this technique are numerous:
shaping of the high harmonic generation spectrum (i.e. quenching or
increasing given harmonic orders), selective excitation of electronic
excited states that are otherwise difficult to reach with conventional
pulses, control of the electronic current in molecular junctions,
etc. Work along these lines is in progress.

This work was partially supported by the Deutsche
Forschungsgemeinschaft within the SFB 658.

%%%%%%%%%%%%%%%%%%%%%%%%%%%%%%%%%%%%%%%%%%%%%%%%%%%%%%%%%%%%%%%%%%%%%%%%%%%%%%%
% Bibliography
%\bibliographystyle{prsty}
%\bibliographystyle{apsrev}
%BibTex database laden mit
\bibliography{paper}
%%%%%%%%%%%%%%%%%%%%%%%%%%%%%%%%%%%%%%%%%%%%%%%%%%%%%%%%%%%%%%%%%%%%%%%%%%%%%%%

\end{document}